\documentclass[prd,aps,preprint,tightenlines,superscriptaddress]{revtex4}
\usepackage{epsfig}
\usepackage{amsmath}
\usepackage{axodraw}
\preprint{DOE/ER/40762-274}
\preprint{UM-PP\#03-035} 

\begin{document}
\title{Counting Rule for Hadronic Light-Cone Wave Functions}
\author{Xiangdong Ji}
\email{xji@physics.umd.edu}
\affiliation{Department of Physics,
University of Maryland,
College Park, Maryland 20742, USA}
\author{Jian-Ping Ma}
\email{majp@itp.ac.cn}
\affiliation{Institute of Theoretical Physics,
Academia Sinica, Beijing, 100080, P. R. China}
\author{Feng Yuan}
\email{fyuan@physics.umd.edu}
\affiliation{Department of Physics,
University of Maryland,
College Park, Maryland 20742, USA}

\date{\today}
\vspace{0.5in}          
\begin{abstract}
We introduce a systematic way to write down the 
Fock components of a hadronic light-cone wave function
with $n$ partons and orbital angular momentum projection $l_z$.
We show that the wave function amplitude $\psi_n(x_i,k_{i\perp},l_{zi})$
has a leading behavior $1/(k^2_\perp)^{[n+|l_z|+{\rm min}(n'+|l_z'|)]/2-1}$ 
when all parton transverse momenta are uniformly large, where 
$n'$ and $l_z'$ are the number of partons and orbital 
angular momentum projection, respectively, of an amplitude that 
mixes under renormalization. The result can be used as a constraint 
in modeling the hadronic light-cone wave functions. We also 
derive a generalized counting rule for hard exclusive processes 
involving parton orbital angular momentum and hadron helicity flip. 

\end{abstract}

\maketitle

Light-cone wave functions are useful tools to describe 
physics of hadrons in high-energy scattering. 
They are snap-shots of hadrons when the latter are moving 
with the speed of light (infinite momentum frame). 
These wave functions can be obtained, in principle, through 
solving the eigen-equation of the light-cone
Hamiltonian using either analytical or numerical methods 
\cite{Brodsky:1998de,Burkardt:1996ct}.
They can also be obtained from the Bethe-Salpeter
amplitudes by integrating out the $k^-$ components
of the parton four-momenta if the latter are known. 
Their moments in momentum space can be 
calculated using lattice QCD or the QCD sum-rule methods 
\cite{Martinelli:1989rr,Chernyak:1984ej}. 
In phenomenological approaches, the light-cone wave
functions are parametrized to fit experiment data 
\cite{Brodsky:1981jv,Chung:1991st,Schlumpf:1993vq,Bolz:1996sw}. 

In Ref. \cite{Burkardt:2002uc}, we have proposed a systematic 
way to enumerate independent amplitudes of a 
light-cone wave function by writing down the matrix
elements of a class of light-cone-correlated quark-gluon operators, 
in much the same way that has been used to construct independent light-cone distribution
amplitudes in which the parton transverse momenta are integrated over
\cite{Braun:1999te}. In Ref. \cite{Ji:2002xn}, we have applied 
this approach to the nucleon, finding that six 
amplitudes are needed to describe the three-quark sector of the 
nucleon wave function. 

In this paper, we introduce a direct method of constructing
the light-cone wave functions in momentum space. By exploiting  
the relations between light-cone amplitudes and the matrix elements
of light-cone-correlated quark-gluon operators, we 
study how the wave function amplitudes depend
on the transverse momenta of partons in the asymptotic limit. 
We find that a general amplitude $\psi_{n}(x_i,k_{\perp i},\lambda_i,l_{zi})$ 
describing an $n$-parton state with orbital angular momentum projection $l_z$ 
goes like
\begin{equation}
    \psi_{n}(x_i,k_{\perp i},\lambda_i,l_{zi}) \rightarrow
  1/(k^2_\perp)^{[n+|l_z|+{\rm min}(n'+|l_z'|)]/2-1}  \ ,
\end{equation}
in the limit that $k_{1\perp}\sim k_{2\perp}\sim ...\sim k_{n-1\perp} \sim k_\perp
\rightarrow \infty$, where $n'$ and $l_z'$ characterize the amplitude
that mixes under scale evolution. The result explains the scaling 
behavior of the $F_2(Q^2)$ form factor obtained recently in perturbative 
QCD \cite{Belitsky:2002kj}, and helps to establish more general scaling
properties of exclusive scattering amplitudes 
\cite{Brodsky:1973kr,Matveev:1973ra,Chernyak:1977as,Lepage:1980fj,Brodsky:1981kj}.
It also can be used as a constraint in building phenomenological 
wave functions of hadrons consistent with perturbative QCD. 


Let us first introduce a systematic method to construct
the light-cone Fock wave function of a hadron with helicity 
$\Lambda$. Suppose a Fock component 
has $n$ partons with creation operators $a^\dagger_1$, 
..., $a^\dagger_{n}$, where the partons can either be gluons or 
quarks and the subscripts label the partons' quantum numbers such as
spin, flavor, color, momentum, etc. Assume all color, 
flavor (for quarks) indices have been coupled properly using 
Clebsch-Gordon coefficients. 
The longitudinal momentum fractions of the partons
are $x_i$ $(i=1,2,...,n)$, satisfying $\sum_{i=1}^n x_i = 1$, 
and the transverse momenta $k_{1\perp},...,k_{n\perp}$, satisfying
$\sum_{i}^n\vec{k}_{i\perp}=0$. We will eliminate $\vec{k}_{n\perp}$ in favor 
of the first $n-1$ transverse momenta. Assume the orbital angular 
momentum projections of the partons are $l_{z1},...,l_{z(n-1)}$, 
respectively, and let $l_z = \sum_{i=1}^{n-1} l_{zi}$, then
\begin{equation}
          l_z + \lambda = \Lambda \ , 
\end{equation}
where $\lambda=\sum_{i=1}^n\lambda_i$ is the total parton helicity.
Without loss of generality, we assume $l_z\ge 0$; even then, $l_{zi}$
can have both signs. Thus, a general term in the hadron wave function 
appears as 
\begin{equation}
     \int \prod_{i=1}^n d[i]~~
        (k_{1\perp}^\pm)^{|l_{z1}|} (k_{2\perp}^\pm)^{|l_{z2}|} 
            ... (k_{(n-1)\perp}^\pm)^{|l_{z(n-1)}|}~ \psi_n(x_i,k_{\perp i},\lambda_i,l_{z i})~
    a_1^\dagger a_2^\dagger ... a_n^\dagger  |0\rangle \ , 
\end{equation}
where $k_i^\pm = k_{ix} \pm k_{iy}$ and the $+(-)$ sign applys
when $l_{zi}$ is positive (negative), and $d[i] = dx_id^2k_{\perp i}/(\sqrt{2x_i}(2\pi)^3)
$ with the overall constraint on $x_i$ and $k_{\perp i}$ implicit.

The above form can be further simplified as follows. Assume $l_{zi}$ is positive 
and $l_{zj}$ negative, and $l_{zi}>|l_{zj}|$, we have
\begin{eqnarray}
   (k_i^+)^{l_{zi}}(k_j^-)^{-l_{zj}}
   &=& (k_i^+)^{l_{zi}+l_{zj}}(k_i^+k_j^-)^{-l_{zj}} \nonumber \\
   &=& (k_i^+)^{l_{zi}+l_{zj}}(k_{i\perp}\cdot k_{j\perp}+i\epsilon^{\alpha\beta}
          k_{i\alpha}k_{j\beta})^{-l_{zj}} \nonumber \\
   &=& (k_i^+)^{l_{zi}+l_{zj}}\left(\phi_0 + \phi_1 
               i\epsilon^{\alpha\beta}
          k_{i\alpha}k_{j\beta}\right) \ , 
\end{eqnarray}
where $\alpha,\beta=1,2$, $\phi_{0,1}$ are polynomials in $k^2_{i\perp}$, 
$k^2_{j\perp}$, and $k_{i\perp}\cdot k_{j\perp}$. On the last line of 
the above equation we have used the identity 
$ \epsilon^{\alpha\beta} \epsilon^{\gamma\delta}
 = \delta^{\alpha\gamma} \delta^{\beta\delta} - \delta^{\alpha\delta} 
       \delta^{\beta\gamma}$. 
If $l_{zi}+l_{zj}\ne 0$, one can use 
$i\epsilon^{\alpha\beta} k_{1\alpha} k_{2\beta} k_1^+
         = k_{1\perp}\cdot k_{1\perp} k_2^+
        - k_{1\perp}\cdot k_{2\perp} k_1^+ $
to further reduce the second term in the bracket.
Following the above procedure, we can eliminate all negative 
$l_{zj}$, a general $l_z>0$ component in the wave function reads
\begin{eqnarray}
   && \int \prod_{i=1}^n d[i]~~
        (k_{1\perp}^+)^{l_{z1}} (k_{2\perp}^+)^{l_{z2}} 
            ... (k_{(n-1)\perp}^+)^{l_{z(n-1)}}~ \nonumber \\ 
 &&  ~~~~ \times \left(\psi_{n}(x_i,k_i,\lambda_i,l_{zi})
          + \sum_{i<j=1|_{ l_{zi}=l_{zj}=0}}^{n-1} i\epsilon^{\alpha\beta} 
              k_{i\alpha}k_{j\beta}\psi_{n(ij)}(x_i,k_{\perp i},\lambda_i,l_{zi})\right)~
    a_1^\dagger a_2^\dagger ... a_n^\dagger  |0\rangle 
\end{eqnarray}
where $\sum_i l_{zi}=l_z$ and $l_{zi}\ge 0$, and the sums over $i$ and 
$j$ are restricted to the $l_{zi}=0$ partons.

Using the above construction, it is easy to see that the proton state with
three valence quarks has six independent scalar amplitudes $\psi^{(i)}_{uud}, i=1,...,6$
\cite{Ji:2002xn}. The wave function amplitudes for three quarks plus one 
gluon will be presented in a separate publication \cite{Ji:2003to}. 

The mass dimension of $\psi_n$ can be determined as follows:
Assume the nucleon state is normalized relativistically
$\langle P|P'\rangle = 2E(2\pi)^3\delta^3(\vec{P}'-\vec{P})$, 
$|P\rangle$ has mass dimension $-1$. Likewise, the parton creation 
operator $a_i^\dagger$ has mass dimension $-1$. Given these, the mass dimension
of $\psi_n$ is $-(n+|l_z|-1)$. The mass dimension of $\psi_{n(ij)}$, however, 
is $-(n+|l_z|+1)$ which can be accounted for by the previous formula with 
an effective angular momentum projection $|l_z|+2$. 


To find the asymptotic behavior of an amplitude $\psi_{n}(x_i,k_i,l_{zi})$
in the limit that all transverse momenta are uniformly large, 
we consider the matrix element of a corresponding quark-gluon 
operator between the QCD vaccum and the hadron state
\begin{equation}
\langle 0|\phi_{\mu_1}(\xi_1)....\phi_{\mu_n}(\xi_n)|P\Lambda\rangle \ , 
\end{equation}
where $\phi$ are parton fields such as the ``good" (+) components of
quark fields or $F^{+\alpha}$ of gluon fields, and
$\mu_i$ are Dirac and transverse coordinate indices when appropriate. 
All spacetime coordinates $\xi_i$ are at equal 
light-cone time, $\xi_i^+=0$. Fourier-transforming with respect to all the spatial 
coordinates ($\xi^-_i$,$\xi_{i\perp}$), we find the matrix element in 
the momentum space, $
\langle 0|\phi_{\mu_1}(k_1)....\phi_{\mu_{n-1}}(k_{n-1})
       \phi_{\mu_n}(0)|p\Lambda\rangle \equiv \psi_{\mu_1,...,\mu_n}(k_1,...,k_{n-1}) $,
here we have just shown $n-1$ parton momenta because of 
the overall momentum conservation. The matrix element
can be written as a sum of terms involving projection
operator $\Gamma^{A}_{\mu_1...\mu_n}(k_{\perp i})$
multiplied by scalar amplitude $\psi_{nA}(x_i,k_{\perp i},l_{zi})$: 
\begin{eqnarray}
\langle 0|\phi_{\mu_1}(k_1)....\phi_{\mu_{n-1}}(k_{n-1})
       \phi_{\mu_n}(0)|p\Lambda\rangle  
  && \equiv  \psi_{\mu_1,...,\mu_n}(k_1,...,k_{n-1}) \nonumber \\
  && = \sum_{A} \Gamma^{A}_{\mu_1...\mu_n}(k_{\perp i})
          \psi_{n}^{(A)}(x_i,k_{\perp i},l_{zi}) \ , 
\end{eqnarray}
where the projection operator $\Gamma^A$ contains 
Dirac matrices and is a polynomial of order $l_z$ 
in parton momenta. For example, the two
quark matrix element of the pion can be written as \cite{Burkardt:2002uc},
\begin{eqnarray}
     &&   \langle 0|\overline{d}_{+\mu}(0)u_{+\nu}(x,k_\perp) 
    |\pi^+(P)\rangle \nonumber \\
   && = (\gamma_5\not\! P)_{\nu\mu} \psi_{u\overline{d}}^{(1)}(x,k_\perp,l_z=0)
            + (\gamma_5\sigma^{-\alpha})_{\nu\mu}P^+ k_{\perp \alpha} 
        \psi_{u\overline{d}}^{(2)}(x,k_\perp,l_z=1) \ ,
\end{eqnarray}
where the projection operators are shown manifestly. More examples for the proton
matrix elements can be found in Ref. \cite{Ji:2002xn}. 

The matrix element of our interest is, in fact, a 
Bethe-Salpeter amplitude projected onto the light cone. 
One can write down formally a Bethe-Salpeter equation 
which includes mixing contributions from other light-cone 
matrix elements.
In the limit of large transverse momentum $k_{\perp i}$, 
the Bethe-Salpeter kernels can be calculated 
in perturbative QCD because of asymptotic freedom. 
In the lowest order, the kernels consist of a minimal 
number of gluon and quark exchanges linking the active 
partons. For the lowest Fock components of the
pion wave function, one gluon exchange is needed to get a large 
transverse momentum for both quarks \cite{Lepage:1980fj}.
As we shall see, asymptotic behavior of the wave function 
amplitudes depends on just the mass dimension of the kernels.

Schematically, we have the following equation for the light-cone
amplitudes,
\begin{eqnarray}
&& \psi_{\alpha_1,...,\alpha_n}(k_1,...,k_{n-1}) 
= \sum_A \Gamma^A_{\alpha_1...\alpha_n}(k_{\perp i}) 
      \psi_{n}^A(x_i, k_{\perp i}, l_{zi}) 
   \nonumber \\
&= & \sum_{n',\beta_1,...,\beta_{n'}}\int d^4q_1 ... d^4q_{n'-1}
       H_{\alpha_1...,\alpha_n,\beta_1,...,\beta_{n'}}(q_i,
      k_i) \psi_{\beta_1,...,\beta_{n'}}(q_1,...,q_{n'-1}) 
\label{bs}
\end{eqnarray}
where $H_{\alpha_1,...,\alpha_n,\beta_1,...,\beta_{n'}}$ are
the Bethe-Salpeter kernels multiplied by the parton propagators. 
When the parton transverse momenta are uniformly large, the kernels
can be approximated by a sum of perturbative diagrams. 
The leading contribution 
to the amplitudes on the left can be obtained by
iterating the above equation, assuming the amplitudes under
the integration sign contain no hard components. As such,
the integrations over $q_{\perp i}$ can be cut-off at a scale $\mu$
where $k_{\perp}>\!\!>\mu>\!\!>\Lambda_{\rm QCD}$, 
and the $q_i$ dependence in $H$ can be expanded in Taylor series. 
In order to produce a contribution to $\psi_n^{(A)}
(x_i,k_{\perp i},l_{zi})$, the hard kernels must contain 
the projection operator $\Gamma^A_{\alpha_1...\alpha_n}(k_1,...,k_{n-1})$. 
Hence we write
\begin{eqnarray}
 && H_{\alpha_1...,\alpha_n,\beta_1,...,\beta_n'}(q_i,k_i)  \nonumber \\
  &=&  \sum_{A,B} \Gamma^A_{\alpha_1...\alpha_n}(k_{\perp i}) 
      H_{AB}(x_i, k_{\perp i}, y_i)
       \Gamma^B_{\beta_1...\beta_{n'}}(q_{\perp i}) \ , 
\end{eqnarray}
where $\Gamma^B_{\beta_1...\beta_{n'}}(q_{\perp i})$ is again a projection 
operator and $H_{AB}(x_i,k_i,y_i)$ are scalar functions of the
transverse momenta $k_{\perp i}$ invariants. Substituting the above into Eq.(\ref{bs}) 
and integrating over $q_i^-$, 
\begin{eqnarray}
&& \psi^{(A)}_n(x_i,k_{\perp i}, l_{zi})
   \nonumber \\
&= & \sum_{B,\beta_i} \int dy_1...dy_{n'-1} H_{AB}(x_i,k_i,y_i) \int d^2q_{\perp 1} ... 
    d^2q_{\perp (n'-1)}
    \Gamma^B_{\beta_1...\beta_{n'}}(q_{\perp i})
          \psi_{\beta_1,...,\beta_{n'}}(y_i,q_i)  \nonumber \\
&= & \sum_{B,\beta_i,A'}\int dy_1...dy_{n'-1} H_{AB}(x_i,k_i,y_i) \int d^2q_{\perp 1} ... 
    d^2q_{\perp (n'-1)}
    \Gamma^B_{\beta_1...\beta_{n'}}(q_{\perp i}) \nonumber \\ && 
      \times  \Gamma^{A'}_{\beta_1...\beta_{n'}}(q_{\perp i})
       \psi_{n'}^{(A')}(y_i,q_{\perp i},l_{zi}') \ ,  
\end{eqnarray}
where the integrations over $q_{\perp i}$ are non-zero
only when the angular momentum content of $\Gamma^B$
and $\Gamma^{A'}$ is the same. Now the large momenta $k_{\perp i}$ 
are entirely isolated in $H_{AB}$ which does not depend on 
any soft scale. The asymptotic behavior of $\psi_n^{(A)}(k_{\perp i})$ is 
determined by the mass dimension of $H_{AB}$, which can be obtained,
in principle, by working through one of the simplest perturbative diagrams.

A much simpler way to proceed is to use light-cone power 
counting in which the longitudinal mass dimension, such as 
$P^+$, can be ignored because of the boost
invariance of the above equation along the $z$ direction.
We just need to focus on the transverse dimensions. 
Since the mass dimension of the amplitudes is $-(n+|l_z|-1)$,
that of $\Gamma^B\Gamma^{A'}$ is $2|l_z'|$, 
and the integration measure $2(n'-1)$, a
balance of the mass dimension yields
$[H_{AB}] = -(n-1+|l_z|)-(n'-1+|l_z'|)$. Therefore, we arrive 
at the central result of the paper that the leading behavior 
of the wave function amplitude goes as
\begin{equation}
   \psi^{(A)}_n(x_i,k_{\perp i}, l_{zi}) \sim \frac{1}
        {(k_\perp^2)^{[n+|l_z|+{\rm min}(n'+|l_z'|)]/2-1}} \ , 
\end{equation}
which is determined by a mixing amplitude with smallest
$n'+|l_z'|$. Since the wave function has mass dimension of 
$-(n+|l_z|-1)$, the coefficient of the asymptotic form must have a soft mass
dimension $\Lambda_{\rm QCD}^{{\rm min}(n'+|l_z'|)-1}$. 

For the quark-antiquark amplitudes of the pion, 
the leading behavior is determined by self-mixing: 
$ \psi_{u\overline{d}}^{(1)}(x,k_\perp,l_z=0) 
~\sim~ 1/k_\perp^2 $ and $\psi_{u\overline{d}}^{(2)}
(x,k_\perp,l_z=1) ~\sim~ 1/(k_\perp^2)^2 $.   
On the other hand, for the three-quark amplitudes of the proton \cite{Ji:2002xn},
we have, $ \psi_{uud}^{(1)}(x_i,k_{\perp i}) ~\sim~ 1/(k_\perp^2)^2$, 
$\psi_{uud}^{(2,3,4,5)}(x_i,k_{\perp i}) ~\sim~ 1/(k_\perp^2)^3,$ 
$\psi_{uud}^{(6)}(x_i,k_{\perp i})~\sim~ 1/(k_\perp^2)^4$. Here
we recall that for $\psi^{(2)}_{uud}$, the effective angular momentum 
projection is $l_z^{\rm eff}=2$. Its leading behavior is determined by its 
mixing with $\psi^{(1)}_{uud}$. 

What are the selection rules for amplitude mixings? First of all, because
of angular momentum conservation, wave function amplitudes 
belonging to different hadron helicity states do not mix. Second, because
of the vector coupling in QCD, the quark helicity in a hard process 
does not change. Therefore, the pion amplitude $\psi^{(2)}_{u\bar d}$ does not 
mix with $\psi^{(1)}_{u\bar d}$ because the total quark helicity differs. 
An example of the nontrivial amplitude mixing is between the pion's 
two-quark-one-gluon and two-quark amplitudes. If one calculates the
asymptotic pion form factor using the wave function amplitudes directly, the
three-parton component does contribute at the leading order. If, however, the 
form factor is calcualted using a factorization approach in which the amplitudes 
are only used at a soft-scale $\Lambda_{\rm QCD}$, the three parton compoent 
contributes as a higher twist.


As an example, we apply the amplitude counting rule 
to hard exclusive processes in which the leading
light-cone wave functions of participating hadrons
dominate. One can, of course, use the light-cone wave functions
to calculate directly hard scattering amplitudes and cross
sections, finding the asymptotic behavior of these
physics observables. Using the expression derived 
for $F_2(Q^2)$ in Ref. \cite{Ji:2002xn} and the asymptotic
behavior of $\psi^{(1)}_{uud}\sim 1/k_\perp^4$ and $\psi^{(3,4)}_{uud}\sim
1/k_\perp^6$, we easily derive
the result found in Ref. \cite{Belitsky:2002kj}:
\begin{equation}
        F_2(Q^2) \sim  1/(Q^2)^{3} \sim F_1(Q^2)/Q^2 \ ,
\end{equation}
in asymptotic limit. On the other hand, the proton amplitudes 
$\psi^{(3,4)}_{uud}$ obtained from Melosh rotation are suppressed
by only one power of $k_\perp$ relative to $\psi^{(1)}_{uud}$, and 
are inconsistent with perturbative QCD in the large $k_\perp$
limit \cite{Melosh:1974cu}. It seems, however, that 
the Melosh-rotated $\psi^{(3,4)}_{uud}$ amplitudes with a harder 
$k_\perp$-dependence are phenomenologically interesting to model 
$F_2(Q^2)/F_1(Q^2) \sim 1/Q$ behavior at an intermediate 
$Q^2$ \cite{Gayou:2001qd,Schlumpf:1993vq,Miller:2002qb,Miller:2002ig}. 

A simpler way to find a generalized counting
rule for hard exclusive processes \cite{Brodsky:1973kr,Matveev:1973ra}
is to count the soft mass dimensions in scattering amplitudes;
the scaling in hard kinematic variables is then determined
by dimensional balance. For example, 
the wave function amplitude $\psi_{n}(x_i,k_i,l_{zi})$ 
when used in a factorization formula 
contains a soft scale factor $\Lambda_{QCD}^{n+|l_z|-1}$. 
Therefore a scattering amplitude involving $H=1,...,N$ hadrons 
with the light-cone amplitudes $\psi_{n_H}(x_i,k_i,l_{zi})$ 
contains a soft mass factor $\Lambda_{QCD}^{\sum_H(
n_H+|l_{zH}|-1)}$. In the hadronic process
$A+B\rightarrow C+D+...$, the fixed-angle scattering
cross section calculated using the amplitudes 
$\psi_{n}(x_i,k_i,l_{zi})$ goes like 
\begin{equation}
  \Delta \sigma \sim    s^{-1 - \sum_H(n_H+|l_{zH}|-1) } \ ,
\end{equation}
where $H$ sums over all hadrons involved.
For $l_{zH}=0$ and minimal $n$, this is 
just the counting rule of Brodsky-Farrar \cite{Brodsky:1973kr} and 
Matveev-Muradian-Tavkhelidze \cite{Matveev:1973ra}. The derivation here emphasizes 
that the traditional counting rule applies only to hadron
helicity conserving processes \cite{Brodsky:1981kj}. The generalized 
counting rule here applies to any hard process proceeded through any 
wave function amplitudes. In particular, it 
reproduces the result of Chernyak and Zhitnitsky for form factors 
where parton orbital angular momentum was first 
considered \cite{Chernyak:1977as}. 

As an application, we consider   
$pp$ elastic scattering. Three helicity conservation amplitudes 
are known to go like $M(++\rightarrow ++) \sim M(+-\rightarrow +-)
\sim M(-+\rightarrow +-)\sim 1/s^4$ \cite{Brodsky:1981kj}. Our counting rule provides 
the scaling behavior of the helicity flipping amplitudes 
$M(++\rightarrow +-)\sim 1/s^{9/2}$ and $M(--\rightarrow ++)\sim 1/s^5$.  

We end the paper with a few cautionary notes. First, we have ignored
the Lanshoff type of contributions in hadron-hadron scattering 
\cite{Landshoff:1974ew}. Second, in an actual calculation
of a scattering amplitude, there are integrations over partons' light-cone 
fractions $x_i$. These integrations may be divergent at the endpoints 
$x_i=0,1$ depending upon the choices of the light-cone wave functions. The 
QCD factorization and the naive power counting break down there 
\cite{Beneke:2002bs,Hoyer:2002qg}.
Finally, the light-cone wave functions defined in the light-cone gauge 
have singularities \cite{Collins:1989bt}. When regularized, 
Sudakov type of form factors appear which lead to the dependence of
the light-cone wave functions on $P^+$ \cite{Botts:1989kf}. 
The $k_\perp$ counting breaks down in the region where the Sudakov 
form factors are important. However, in certain cases the endpoint singularities 
are regulated by the Sudakov effects, and the last two adverse factors 
cancel \cite{Li:1992nu}, leaving the naive counting rule intact. It 
is not clear, however, that this happens in general.

X. J. and F. Y. were supported by the U. S. Department of Energy via 
grant DE-FG02-93ER-40762. J.P.M. was supported by National Natural 
Science Foundation of P.R. China through grand No.19925520. 

\bibliography{protonref}

\end{document}